# LITERATURE REVIEW OF ATTRIBUTE LEVEL AND STRUCTURE LEVEL DATA LINKAGE TECHNIQUES

Mohammed Gollapalli

College of Computer Science & Information Technology,
University of Dammam, Dammam, Kingdom of Saudi Arabia

## ABSTRACT

*Data Linkage is an important step that can provide valuable insights for evidence-based decision making, especially for crucial events. Performing sensible queries across heterogeneous databases containing millions of records is a complex task that requires a complete understanding of each contributing database's schema to define the structure of its information. The key aim is to approximate the structure and content of the induced data into a concise synopsis in order to extract and link meaningful data-driven facts. We identify such problems as four major research issues in Data Linkage: associated costs in pair-wise matching, record matching overheads, semantic flow of information restrictions, and single order classification limitations. In this paper, we give a literature review of research in Data Linkage. The purpose for this review is to establish a basic understanding of Data Linkage, and to discuss the background in the Data Linkage research domain. Particularly, we focus on the literature related to the recent advancements in Approximate Matching algorithms at Attribute Level and Structure Level. Their efficiency, functionality and limitations are critically analysed and open-ended problems have been exposed.*

## 1. INTRODUCTION

Organizations worldwide have been collecting data for decades. Data collected from The World Bank [24], The National Climatic Data Centre [49], and countless other private and public organizations have been collecting, storing, processing and analysing massive amounts of data which has the potential to be linked for the discovery of underlying factors to critical problems. Sharing of large databases between organizations is also of growing importance in many data mining projects, as data from various sources often has to be linked and aggregated in order to improve data quality, or to enrich existing data with additional information [7]. When integrating data from different sources to implement a data warehouse, organizations become aware of potential systematic differences, limitations, restrictions or conflicts which fall under the umbrella-term data heterogeneity [34]. Poor quality data has also been prevalent in databases due to a variety of reasons, including typographical errors, lack of standards etc. To be able to query and integrate data in the presence of such data uncertainties as depicted in Fig. 1, a central problem is the ability to identify whether heterogeneous database tables, attributes and tuples can be linked with the primary aim to understand the past and predict the future.

In response to the aforementioned challenges, significant advances have been made in recent years in mining structures of databases with the aim to acquire crucial fact finding information that is not otherwise available, or that would require time-consuming and expensive manual procedures. Schemas are definitions that identify the structure of induced data and are the result of a database design segments. The relational database schemas that are invariant in time hold valuable information in their tables, attributes and tuples which can aid in identifying semantically similar objects. The process of identifying these schema structures has been one of the essential elements of data mining process [21-26]. Accurate integration of heterogeneous database schema can provide valuable insights that are useful for evidence-based decision making, especially for crucial events. In the schema integration process, each individual database can be analysed to provide and extract local schema definitions of the data. These local schema definitions can be used for the development of a global schema which integrates and subsumes the local schema in such a way that (global) users are provided with a uniform and correct view of the global database [19]. With the help of global schema structures, we can derive hierarchical relationships up to the instance level across datasets. However, without having this global schema, extracting meaningful data into a usable form can become a tedious process [5, 8, 14, 18, 21, and 26]. Traditional local-to-global schema-based techniques also lack the ability to allow computational linkage and are not suitable when dealing with heterogeneous databases [2, 5, 8, 18, 57, 61 and 66]. To make things worse, the data could be "dirty" and differences might exist in the structure and semantics maintained across different databases. Research communities have also stressed Schema Pattern Matching [21 to 26] and SQL Querying [27, 28]. Schema Pattern Matching uses database schema to devise clues as to the semantic meaning of the data. Constraints are used to define requirements, generated by hand or through a variety of tools. However, the main problems with Schema Pattern Matching are insufficiency and redundancy.

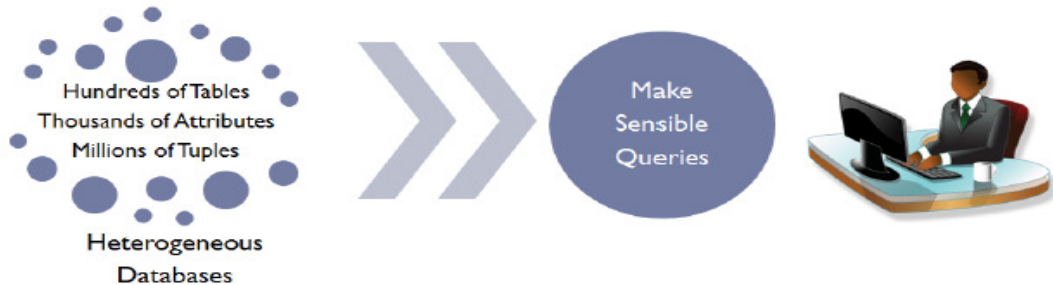

Figure 1. Data linkage across heterogeneous databases

Data linkage (also known as data matching, probabilistic matching, and instance identification) is the process of identifying records which represent the same real world entity despite typographical and formatting constraints [18, 25, 32, 34, and 37]. In conducting our research, we observed four prime areas where data linkage is a persistent, yet heavily researched problem.

1. Medical science for DNA sequence matching and biological sequence alignment [12, 18, 21, 47, 56, and 80-84].
2. Government departments for taxation and pay-out tracking [5, 24, 30, 48, and 79].

3. Businesses integrating the data of acquired companies into their centralized systems [2, 36, and 42].

4. Law enforcement for data matching across domains, such as banking and the electoral commission [24, 30, 33, 49, and 50].

Traditional data linkage approaches use similarity scores that compare tuple values from different attributes, and declare it as matches if the score is above a certain threshold [2, 10, 18, 61, 67, and 79]. These approaches perform quite well when comparing similar databases with clean data. However, when dealing with a large amount of variable data, comparison of tuple values alone is not enough [1, 2]. It is necessary to apply domain knowledge when attempting to perform data linkage where there are inconsistencies in the data. The same problem applies to database migrations, and to other data intensive tasks that involve disparate databases without common schemas. Furthermore, the creation of data linkage between heterogeneous databases requires the discovery of all possible primary and foreign key relationships that may exist between different attribute pairs, on a global spectrum [1, 3, 8, 11, and14-16].

## 2. TAXONOMY OF DATA LINKAGE APPROACHES

Different techniques have been presented by researchers [18, 32, 34, 35, 43, and 77] in multiple areas which argue that the need, task, and type of linkage to be performed will define the involved steps. Other techniques such as the Statistic New Zealand [48] lean toward the idea that data linkage will always require manual preliminary steps such as data classification, sampling and missing observation detection. However, the fundamental problem that arises each time in performing data linkage on large volumes of heterogeneous databases is to discover all possible relationships based on matching similar tuple values that might exist between different table attributes [1].

In this paper, we review on techniques that exist in performing approximate data linkage based on their approach rationale. We compare the advantages and disadvantages of current approaches for solving data linkage problem in multiple ways. Our analysis of existing techniques as depicted in Fig. 2 will show that there is room for substantial improvement within the current state-of-the-art and we recommend techniques where further improvements can be made.

### 2.1. SQL Matching Strategies

SQL Matching techniques [14, 21, 22, 23, 25 and 26] perform data linkage using simple SQL-LIKE commands and SQL Extensions. The advantage of SQL matching techniques is that they help in performing quick data linkage across databases. However, they do not perform well in cases where comparison and identification of data structures need to be performed on large databases containing noisy data without proper unique keys, foreign key relationships, indexes, constraints, triggers, or statistics. Another drawback of the SQL matching process is that it performs |m| x |n| time's column match where m and n are the total tuple counts in two different databases, resulting in a very slow, expensive and tedious process.

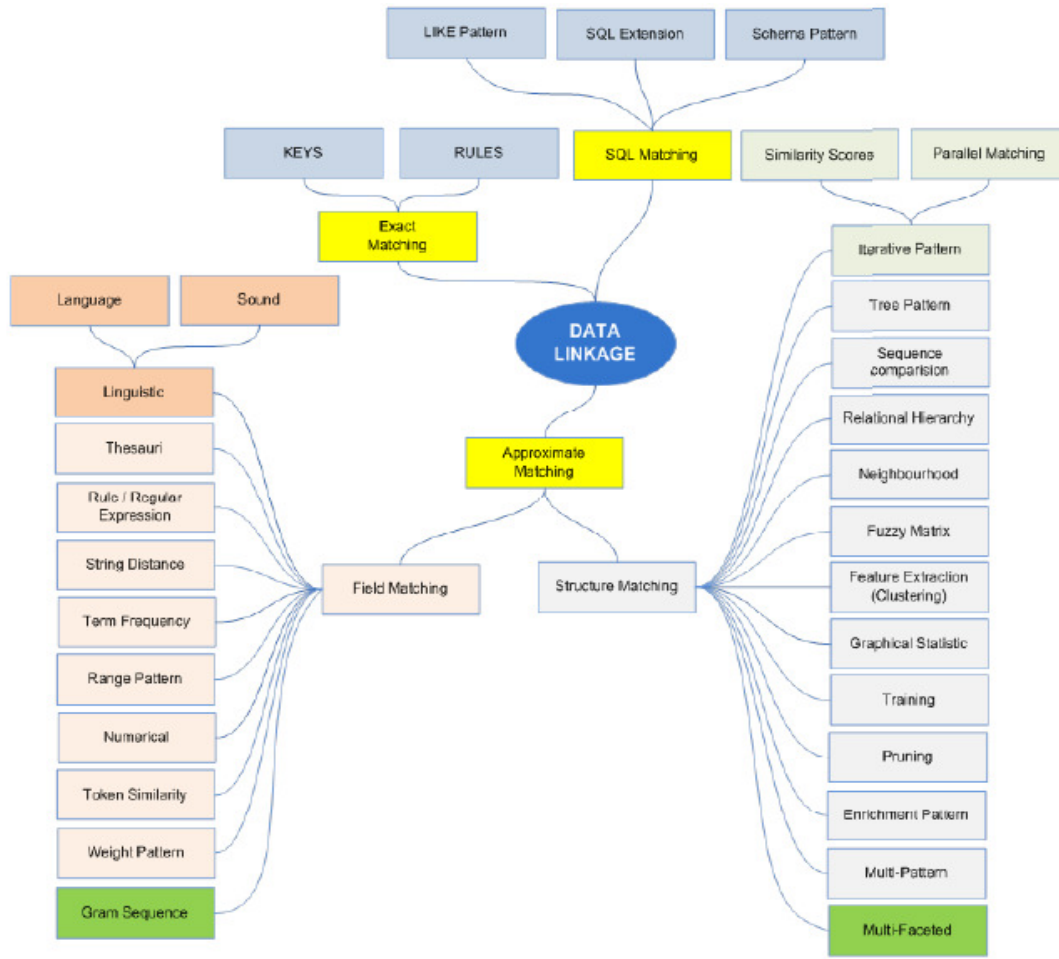

Figure 2. Data linkage approaches

A variation of SQL Matching includes extending query syntax functionalities to perform data linkage. The proposed SQL-LIKE Command languages [22, 23 and 26] handle data transformation, duplicate elimination and cleaning processes supported by regular SQL Query and a proposed execution engine. However, these techniques demand users to have significantly advanced SQL scripting skills and proposed extended functionalities along with sound domain knowledge. Thus, syntax based SQL matching techniques are proven to be less attractive in real world scenarios [22].

Research communities have also stressed Schema Pattern Matching [21 to 26] and SQL Querying [27, 28]. Schema Pattern Matching uses database schema to devise clues as to the semantic meaning of the data. Constraints are used to define requirements, generated by hand or through a variety of tools. However, the main problems with Schema Pattern Matching are insufficiency and redundancy. SQL Querying, on the other hand, uses a SQL query language such as the Resource Description Framework (RDF) [27, 28] to define matching criteria. Difficulties arise when restrictions eliminate the discovery of possible matches. More relaxed queries use a structure-free mechanism by applying a tree pattern query; however, tree-pattern queries are highly inaccurate due to a high incidence incorrect manual identification of relationships [29].

## 2.2. Exact Matching Strategies

Unlike SQL Matching, Exact Matching techniques give more insight into the content and meaning of schema elements [25]. Exact matching uses a unique identifier present in both datasets being compared. The unique identifier can only be linked to one individual item, or an event (for example, a driver's license number). The Exact Matching technique is helpful in situations where the data linkage to be performed belongs to one data source. For example, consider a company with a recent system crash willing to perform data linkage between the production data source file and the most recent tape backup file to trace transactions. In such situations, Exact Matching would likely suffice in performing data linkage. A specific variation of exact matching discovered In this research is the Squirrel System [31], using a declarative specification language, ISL, to specific matching criteria which will match one record in a given table, with one record in another table. However, exact matching approach leaves no room for uncertainty; records are either classified as a match or as a non-match. Problems often arise when the quality of the variables does not sufficiently guarantee the unique identifier is valid [16]. Exact matching comparison does not suffice for matching records when the data contains errors, for example typographical mistakes, or when the data have multiple representations, such as through the use of abbreviations or synonyms [10].

## 2.3. Approximate Matching Strategies

Approximate matching is also known as the probabilistic approach [34 to 36] within the research community and is highly recommended, state-of-the art, alternative approach compared to exact matching. In approximate matching techniques, data linkage is performed on a likelihood basis (i.e. performing matching based on the success threshold ratio). Output results can vary in different formats such as "match, possible match, and non-match" basis, Boolean type true or false match basis, nearest and outermost distance match basis, discrete or continuous match basis etc. Variations in approximate matching technique include statistical and probabilistic solutions for similarity matching. Attention has also been drawn to approximate matching techniques from different research arenas, including statistical mathematics and bio-medical sciences.

Due to the variety of proposed approaches and the level of attributes match, we have focused our research and classified most common approximate matching techniques into attribute level matching and structure level matching groups discussed in the next two sections. It is important to note that, the purpose of this paper is not to list every data linkage technique rather to discuss the multitude of approximate matching techniques available in the areas of attributes and structure level matching. At the end of this paper, we discussed our conclusions and recommendations for future work.

## 3. ATTRIBUTE LEVEL MATCHING

Attribute Matching, also known as Field Matching [35] and Static String Similarity [36] deals with one-to-one match across different data sources. A challenging task of attribute matching is to perform data linkage across data sources by comparing similar matching records with the assumption that the user is aware of the database structure. Individual record fields are often stored as strings, meaning that functions which accurately measure the similarity of two strings are important for deduplication [36]. In the following subsections, we describe most commonly used attribute matching methodologies and discuss their efficiency.

## 3.1. Linguistic similarity

Linguistic techniques focus on phonetic similarities between strings. The rationale behind this approach is that while strings may be similar phonetically, they may have different characters to locate potential matches. Soundex [34] is the most widely known in this area, and uses codes to define letters, remaining non-coded letters are used as separators. In addition, Soundex checks for identical codes (A, E, I, O, U and Y) without separators. Through the Soundex rules, a possible match is determined or denied. Advantages of linguistic techniques include the exposure of about 2/3 of spelling variations [25, 32, and 34]. However, linguistic methods are not equally effective from one ethnicity to the next. Linguistic based techniques are designed for Caucasians, and works on most other ethnicities, but largely fails on East Asian names due to the phonetic differences. NYSIIS [34] improved upon this by maintaining vowel placement and converting all vowels to the letter A. Nonetheless, it is still not perfectly accurate and performs best on surnames and not on other types of data [34].

## 3.2. Rule/Regular expression

The Rule / Regular expression [40] approach uses rules or set of predefined regular expressions and perform matching on tuples. Regular Expression Pattern as proposed in [40] is more flexible than regular expression alone, which is built from alphabetical elements. This is also because the Regular Expression Pattern is built from patterns over a data element, allowing the use of constructs such as "wildcards" or pattern variables. Regular Expression Pattern is quite useful when manipulating strings, and can be used in conjunction with basic pattern matching. However, the problem with this approach lies in the fact that it is relatively domain specific and tends to only work well on strings.

## 3.3. Ranking

Ranking [15, 41] methods determine preferential relationships and have been more recently recognized by researchers as a necessary addition to structure based matching techniques. Search engines have used ranking methods for some time, such as Google's PageRank, despite such algorithms not suited for matching noisy data due to their poor connectivity and lack of referrals [15]. Therefore, ranking extensions which simultaneously calculate meaning and relevance are researched. Thus far, only a few ranking methods have been proposed including induction logic programming, probabilistic relational kernel, and complex objects ranking [15, 41].

## 3.4. String distance

String distance methods, also known as character-based similarity metrics [34] are used to perform data linkage based on the cost associated within the comparing strings. The cost is estimated on the number of characters which needs to be inserted, replaced or deleted for a possible string match. For example, Fig. 3 shows the cost associated in editing string "Aussie" to "Australian" (the "+" sign shows addition, the "-" sign shows deletion, and the "x" sign shows replacement).

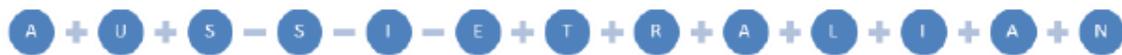

Figure 3. An example of string distance technique

Experimental results in [34] have shown that the different distance based methodologies discovered so far are efficient under different circumstances. Some of the commonly recommended distance based metrics include Levenstein distance, Needleman-Wunsch distance, Smith-Waterman distance, Affine-gap distance, Jaro metric, Jaro and Jaro-Winkler metric, Q-gram distance, and positional Q-grams distance. Through the various methods, costs are assigned to compensate for pitfalls in the system. Yet, overall, string distance pattern is most effective for typographical errors, but is hardly useful outside of this area [34].

## 3.5. Term frequency

Term frequency [43] approach determines the frequency of strings in relation and to favour matches of less common strings, and penalizes more common strings. The Term frequency methods allow for more commonly used strings to be left out of the similarity equation. TF-IDF [43] (Term Frequency-Inverse Document Frequency) is a method using the commonality of the term (TF) along with the overall importance of the term (IDF). TF-IDF is commonly used in conjunction with cosine similarity in the vector space model. Soft TF-IDG [44] adds similar token pairs to the cosine similarity computation. According to the researchers in [44], TF-IDF can be useful for similarity computations due to its ability to give proportionate token weights. However, this approach fails to make distinctions between the similarity level of two records with the same token or weight, and is essentially unable to determine which record is more relevant.

## 3.6. Range pattern

Range pattern matching returns a Boolean style true or false result if the specified tuples fall within the specified range. Similarity or dissimilarity is determined when the elements of the data are compared against the predetermined range. Range matching will return a 0 or 1, with 0 being false and 1 being true. Range pattern matching is often used as an expansion of an algorithm to filter results. For example, TeenyLIME [45] expands upon LIME by adding range pattern capabilities, giving TeenyLIME the ability to define the range of its results. A drawback of the range pattern approach is that it is often not powerful enough to perform matching without a high level of query knowledge. For example, if a query is made to search for nearby locations, an optimal range is often not given or is defined by words having various meanings, causing range pattern matching to produce inaccurate results.

## 3.7. Numeric distance

Numeric distance methods are used to quickly perform data linkage on tuples that contains numerical values but don't require complex string character-style comparison. Hamming distance [46], for example, is used for numeric values such as zip codes, and counts the variations between two records. Due to the limitations of numeric data type constraints, it has not received much attention. Numeric distance methods can be best used in combination of other techniques.

## 3.8. Token matching

Token based matching compare fields by ignoring the ordering of the tokens (words) within these fields. Token based approach use tokenization to perform matching, which is the separation of strings into a series of tokens. It assigns a token to each word in the string and tries to perform matching by ignoring token order and by performing similar match. The token based approach attempts to compensate for the inadequacies of character-based metrics, specifically the inability to detect word order arrangement. A tokenizer performs the operation, taking into account characters, punctuation marks, blank spaces, numbers, and capitalisation. Token based methods count a string as a word set, and accommodates duplicates. For example, Cosine Similarity [38] is used to perform data linkage based on record strings, irrespective of word ordering within the string. The Cosine Similarity methods are effective over a range of entry types, and also have the advantage of considering word location to allow for swapping of word positions. For data containing a large amount of text, the token based matching works quite well, as it can handle repeating words. The optimising token based approach has typically included aggregation of different sources. A potential drawback is that token based matching does not store sub-string order and can predict false matches.

## 3.9. Weight pattern

Weight pattern also referred to as Scoring [47], is applied on matching strings to return a numerical weight; a positive weight for agreeing values and a negative weight for disagreeing values. As two records are compared, the system assigns a weight value for similarity comparison. Composite weight [48] is a summation of all the field weights for a record, which multiplies the probabilities of each value. Reliability of the information, commonality of the values, and similarity between the values are considered in determining weight. Determinations are made by calculating the "m" probability (reliability of data) and the "u" probability (the commonness of the data). For example, IDF weights consider how often a particular value is used. After weights are determined for all the data, cut-off thresholds are set to determine the comparison range. Unfortunately, weight pattern techniques do not perform well when there are data inconsistencies. True matches may have low weights, and non-matches may have high weights as a result of simple data errors [48].

## 3.10. Gram sequence

Gram sequence based techniques compare the sequence of grams of one string with the sequence of grams of another string. n-grams is a gram based comparison function which calculates the common characters in a sequence, but is only effective for strings that have a small number of missing characters [46]. For example, the strings "Uni" and "University" have the same 2-gram {un, ni}. q-gram [85] involves generating short substrings of length q using a sliding window at the beginning and end of a string [85]. The q-gram method can be used in corporate databases without making any significant changes to the database itself [85]. Theoretically, two similar strings will share multiple q-grams. Positional q-grams record the position of q-grams within the string [14]. Danish and Ahy in [85] proposed to generate q-grams along with various processing methods such as substrings, joins, and distance. Unfortunately, the gram sequence approach is only efficient for short string comparison and becomes complex, expensive and unfeasible for large strings [85].

## 3.11. Blocking

Blocking [46] techniques separate tuple values into set of blocks/groups. Within each of these blocks, comparisons are made. Sorted Neighborhood is a blocking method which first sorts and then slides a "window" over the data to make comparisons [46]. BigMatch [51] used by the U.S. Census Bureau, is another blocking technique. BigMatch identifies pairs for further processing through a more sophisticated means. The blocking function assigns a category for each record and identical records are given the same category. The disadvantage of the blocking method is that it will not work for records which have not been given the same category [18, 25, and 34].

## 3.12. Hashing

Hashing methods convert attributes into a sequence of hash values which are compared for similarity matching between different sets of strings. Hashing methods require conversion of all the data to find the smallest hash value, which could be a costly approach. Set-of-sets [8] is a hashing based data matching technique which works reasonably well in smaller string matching scenarios. The set-of-sets technique proposed in [8] divides strings into 3-grams and assigns a hash value to each tri-gram. Once hash values are assigned and placed in a hash bag, only the lowest matching hash values are considered for matching. Unfortunately, this technique doesn't yield accurate results when dealing with variable length strings and uses traditional hashing which results in completely different hash values for even a small variation [79]. Furthermore, the Set-of-sets requires conversion of all the data prior to comparison in order to find the smallest hash value, which could be a costly approach. To overcome this disadvantage, the h-gram (hash gram) method was proposed in [79] to address the deficits of the set-of-sets technique, by extending the n-gram technique; utilizing scale based hashing; increasing matching probability; and by reducing the cost associated in storage of hash codes.

## 3.13. Path sequence

The path sequence approach such as in [37] examines the label sequences, and compares them to the labelled data. The distance is measured by determining the similarity between the last elements of a path. The prefix can be considered, but this only affects the result to a certain degree, and becomes less relevant with increasing distance between the prefix and the end of the sequence.

## 3.14. Conditional substrings

Substring matching such as in [53] expands upon string-based techniques by adding substring conditions to string algorithms. Distance measurements are calculated for the specified substring, in which all substring elements must satisfy the distance threshold. A frequent complication related to conditional substring based matching involves the estimation of the size of intersection among related substrings. Clusters and q-grams [2, 4, and 53], which are commonly used in string estimation, are not applicable in substring based techniques, because substring elements are often dissimilar. As a result, substring matching is hindered by an abundance of possibilities, which must all be considered.

## 3.15. Fuzzy Matrix

Fuzzy Matrix [32, 60] places records in the form of matrices and apply fuzzy matching techniques to perform record matching. Commonly used by social scientists to analyse behavioural data, the fuzzy matrix technique is also applicable to many other data types. When considering a fuzzy set, a match is not directly identified as positive or negative. Instead, the match is considered on its degree level of agreement with the relevant data. As a result, a spectrum is created which identifies all levels of agreement or truth.

### 3.16. Thesauri Matching

Thesauri based matching attempts to integrate two or more thesauruses. A thesaurus is a kind of lexicon to which some relational information has been added, containing hyponyms which give more specific conceptual meaning. WordNet [27, 32, and 52] is a public domain lexical database, or thesaurus, which makes its distinctions by grouping words into sets of synonyms; it is often used in thesauri matching techniques. Falcon and DSSim [52] are thesauri based matching tools which incorporate lexicons, edit-distance and data structures. LOM [32] is a lexicon-based mapping technique using four methods (whole term, word constituent, synset, and type matching) in an attempt to reduce the required amount of human labour, but does not guarantee any level of accuracy. While Thesauri based approaches can be extremely useful in merging conceptual, highly descriptive information; they can be incredibly complex and difficult to automate to a significant degree; and human experts are typically required to quality assure the relationships [27]. Thesauri matching algorithms also needs to consider the best balance between precision and recall.

## 4. STRUCTURE LEVEL MATCHING

Structure level matching is used when the records being matched need to be fetched from a combination of records (i.e. when attempting to match noisy tuples across different domains, and requiring more than one match). These techniques perform data matching, with the main intuition that the grouping of attributes into clusters followed by performing matching provides a deeper analysis of related content and semantic structure. This process was initially considered for discovering candidate keys and dependent keys. However, one of the biggest challenges involved in this process has been the large number of combinations required for grouping attributes and performing data matching between these groups, which can be costly and time consuming [25, 32, 34 and 37]. Large scale organisations such as Microsoft and IBM have introduced Performance Tuner tools for indexing combined attributes on which queries are frequently executed. Unfortunately, these tools are suited to Database Developers / DBA's who have sound knowledge in executing SQL queries and is not ideal for novice users. As such, research has taken new directions by classifying multiple structure level techniques that require matching across multiple attributes. We have classified principal techniques in the following subsections.

### 4.1. Iterative pattern

Iterative pattern is the process of repeating a step multiple times (or making "passes") until a match is found based on similarity scores and blocking variables (variables set to be ignored for similarity comparison). The Iterative approach uses attribute similarity, while considering the similarity between currently linked objects. For example, the Iterative pattern method will consider a match of "John Doe" and "Jonathan Doe" as a higher probability if there is additional matching information between the two records (such as spouse's name and children's names). The first part of the process is to measure string distance, followed by a clustering process. Iterative pattern methods have proven to detect duplicates that would have likely been missed by other methods [54]. The gains are greater when the mean size of the group is larger, and smaller when the mean size is smaller. Disadvantages surface when distinctive cliques do not exist for the entities or if references for each group appear randomly. Additionally, there is also the disadvantage of cost, as the Iterative pattern method is computationally quite expensive [54].

## 4.2. Tree pattern

Tree pattern is based on decision trees with ordered branches and leaves. The nodes are compared based on the extracted tree information. CART and C.5 are two widely-known decision tree methods which create trees through an extensive search of the available variables and splitting values [55]. A Tree pattern starts at the root node and recursively partitions the records into each node of the tree and creates a child to represent each partition. The process of splitting into partitions is determined by the values of some attributes, known as splitting attributes, which are chosen based on various criteria. The algorithm stops when there are no further splits to be made. Hierarchical verification through trees examines the parent once a matching leaf is identified. If no match is found within the parent, the process stops; otherwise the algorithm continues to examine the grandparent and further up the tree [37]. Suffix trees such as DAWG [37] build the tree structure over the suffixes of S, with each leaf representing one suffix and each internal node representing one unique substring of S. DAWG has additional feature of failure links added in for those letters which are not in the tree. Disadvantages of Tree pattern lies in lengthy and time consuming process with manual criteria often needed for splitting.

## 4.3. Sequence pattern

Sequence pattern methods perform data linkage based on sequence alignment. This technique attempts to simulate a sequential alignment algorithm, such as the BLAST (Basic Local Alignment Search Tool) [12] technique used in Biology. The researchers compared the data linkage problem with the gene sequence alignment problem for pattern matching, with the main motivation to use already invented BLAST tools and techniques. The algorithm translates record string data into DNA sequences, while considering the relative importance of tokens in the string data [12].

Further research in the Sequence pattern area have exposed variations based on the type of translation used to translate strings into DNA Sequence (i.e. weighted, hybrid, and multi-bit BLASTed linkage) [12]. BLASTed linkage has advantages through the careful selection of one of its four variations, as each variation performs well on specific types of data. Unfortunately, sequence pattern tends to perform poorly on particular data strings, depending upon the error rate, importance weight, and number of common tokens [12].

## 4.4. Neighbourhood pattern

The neighbourhood approach [7, 59] attempts to understand and measure distribution according to their pattern match, and is a primary component in identifying statistical patterns. By using the nearest neighbour approach, related data is able to be clustered even if it is specifically separated. The logic behind this approach is based on the assumption that, if clustered objects are similar, then the neighbours of clustered objects have a higher likelihood of also being similar. Neighbourhood pattern requires a number of factors that need to be carefully considered in order to determine pattern matches which is considered as a key downfall.

## 4.5. Relational hierarchy

Relational Hierarchy techniques use primary and foreign key relationships to understand related table content in order to perform data linkage. Relational hierarchy forms relation links which connect concepts within various categories. It breaks down the hierarchical structure and the top-level structure contains children sets. The relational hierarchy technique compares and calculates the co-occurrence between tuples by measuring the overlap of the children sets. A high degree of overlap will indicate a possible relationship between the two top level categories [57]. Relational Hierarchy techniques are only effective when primary and foreign key relationships have been established. Raw data, without predefined relationships, cannot be linked using this approach.

## 4.6. Clustering/Feature extraction

Clustering, also known as the Feature extraction method performs data linkage based on common matching criteria in clusters, so that objects in clusters are similar. Soft clustering [61], or probabilistic clustering, is a relaxed version of clustering which uses partial assignment of a cluster centre. The SWOOSH [62] algorithms apply ICAR properties (idempotence, commutativity, associativity, representativity) to the match and merge function. With these properties and several assumptions, researchers introduced the brute force algorithm (BFA), including the G, R and F SWOOSH algorithms [44]. SIMCLUST is another similarity based clustering algorithm which places each table in its own cluster as a starting point and then works its way through all of the tables by consecutively choosing two tables (clusters) with the highest level of similarities. [5] proposed iDisc system which creates database representations through a multi-process learning technique. Base clusters are used to uncover topical clusters which are then aggregated through meta-clustering. Clustering in general can get extremely complex (such as forming clusters using semantics) and needs to be handled carefully while discovering relationships between matching clusters.

## 4.7. Graphical statistic

Graphical statistic is a semi-automated analysis based technique where data linkage is performed based on the results obtained on the graph. Such representations illustrate the topical database structure through tables. The referential relationship indicates an important linkage between two separate tables. Foreign keys within one table may refer to keys within the second table. However, problems with this technique often arise due to the fact that information on foreign keys is often missing [5].

## 4.8. Training based

Training based technique is a manual approach where users are constantly involved in providing statistical data based on previous/future predictions. In [7], researchers presented a two-step training approach using automatically selected, high quality examples which are then used to train a support vector machine classifier. The approach proposed in [7] outperforms k-means clustering, as well as other unsupervised methods. The Hidden Markov training model, or HMM, standardises name and address data as an alternative method to rule-based matching. Through use of lexicon-based tokenization and probabilistic hidden Markov models, the approach attempts to cut down on the heavy computing investment required by rule programming [64]. Once trained, the HMM can determine which sequence of hidden states is most likely to have emitted the observed sequence of symbols. When this is identified, the hidden states can be associated with words from the original input string. This approach seems advantageous in that it cuts down on time costs when compared to rule-based systems. However, this approach remains a lengthy process, and has shown to run into significant problems in various areas. For instance, HMM confuses given, middle, and surnames, especially when applied to homogenous data. Furthermore, outcomes proved to be less accurate than those of rule-based systems [64]. DATAMOLD [65] is a training-based method which enhances HMM. The program is seeded with a set of training examples which allows the system to extract data matches. A common problem with training techniques is that it requires many examples to be effective; and the system will not perform without an adequate training set [55].

## 4.9. Pruning/Filtering statistic

Pruning statistic performs data linkage by trimming similar records on a top down approach. In [16], the data cleaning process of "deduplication" involves detecting and eliminating duplicate records to reduce confusion in the matching process. For data which accepts a large number of duplicates, pruning, before data matching, simplifies the process and makes it more effective. A pruning technique proposed by Verykios [34] recommends pruning as on derived decision trees used for classification of matched or mismatched pairs. The pruning function reduces the size of the trees, improving accuracy and speed [34]. The pruning phase of CORDS [16] (which is further discussed in the statistical analysis section) prunes non-candidates on the basis of data type, properties, pairing rules, and workload; such tasks are done to reduce the search space and make the process faster for large datasets. Pruning techniques [37] are based on the idea that it is much faster to determine non-matching records than matching records, and therefore aim to eliminate all non-matching records which do not contain errors. However, the disadvantage of such techniques is that they are not suitable in identifying matches of any type, and must be combined with another matching technique.

## 4.10. Enrichment pattern

Enrichment patterns are a continuous improvement based technique which performs data linkage by enriching the similarity tasks on a case by case basis. An example of the enrichment method is ALIAS [34], a learning-based system, designed to reduce the required amount of training material through the use of a "reject region". Only pairs with a high level of uncertainty require labels. A method similar to ALIAS is created using decision trees to teach rule matching in [34]. OMEN [32] enriches data quality through the use of a Bayesian Net, which uses a rule set to show related mappings. Semantic Enrichment [66] is the annotation of text within a document by sematic metadata, essentially allowing free text to be converted into a knowledge database through data extraction and data linking. Conversion to a knowledge database can be through exact matching or by building hierarchical classifications of terms; text mining techniques allow annotation of concepts within documents which are subsequently linked to additional databases. Thesauri alignment [32, 52] based techniques are also considered as part of enrichment techniques because it combines concepts and better defines the data. The problems associated with enrichment approach include substantial investment of time and the requirement for extensive domain knowledge.

## 4.11. Multi pattern

The multi (multiple) pattern approach performs data linkage through the simultaneous usage of different matching techniques. This approach best fits when one does not know which technique performs better. The researchers in [31] use a multi approach which combines sequence matching, merging, and then exact matching. Febrl [67] is an open-source software containing comparison, and record pair classifications. Febrl results are conveniently presented in a graphical user interface which allows the user to experiment with numerous other methods [67]. TAILOR [46] is another example which uses three different methods to classify records: decision tree induction, unsupervised k-means clustering, and a hybrid approach. GLUE [68] is yet another matching technique allowing for multiple matching methods. GLUE performs matching by first identifying the most similar concepts. Once these concepts are identified, a multi-strategy learning approach allows user to choose from several similarity measures to perform the measurement.

## 4.12. Data constraints

Data constraints, also known as internal structure based techniques, apply a data constraint filter to identify possible matches [43]. The constraint typically uses specific criteria of the data properties. This technique is not suited when used on its own, and performs best for the elimination of non-matches, as a pre-processing method before a secondary method, such as clustering. Furthermore, data constraints don't handle the large number of uncertainties present within the data. Hence, adding constraints for each uncertainty is computationally infeasible.

## 4.13. Taxonomy

Taxonomy based methods use taxonomies, a core aspect of structural concepts which are largely used in file systems and in knowledge repositories [69]. This approach uses the nodes of taxonomy to define a parent/child relationship within the conceptual information and create classification. Using specified data constraints, the taxonomy of multiple data sources are evaluated into a technique known as structural similarity measure. For example, in [70] researchers used a taxonomy mapping strategy to enrich WordNet with a large number of instances from Wikipedia, essentially merging the conceptual information from the two sources. As with similar methods, taxonomy based matching requires a significant degree of domain knowledge and performs with limited precision and inadequate recall.

## 4.14. Hybrid match

Hybrid techniques use a combination of several mapping methods to perform data match. A prime example of the hybrid method is described in [71], which uses a combination of syntactic and semantic comparisons. The rationale behind hybrid matching is that the semantics alone is not sufficient to perform accurate matching and could be inconsistent. The hybrid solution consists of a hybrid of semantic and syntactic matching algorithms which considers individual components. The syntactic match uses a similarity score based on class, prefix and substring, and the semantic match uses a similarity score based on cognitive measures such as LSA, Gloss Vector, and WordNet Vector. The information is aggregated and entered into a matrix and experts are used to determine domains within the selected threshold.

## 4.15. Data extraction

Data extraction primarily involves extracting semantic data. Data extraction can be performed manually or with an induction and automatic extraction [72]. In [73], researchers used data recognisers to perform data extraction on the semantics of data. The recogniser method is aimed at reducing alignment after extraction, speeding up the extraction process, reusing existing knowledge, and cutting down on manual structure creation. This approach is found to be effective for simple unified domains, but not for complicated, loosely unified domains. Another benefit of the data extraction technique is that, after the data is extracted, it can be handled as instances in a traditional database. However, it generally requires a carefully constructed extraction plan by an expert in that specific knowledge domain [74].

## 4.16. Knowledge integration

Knowledge integration techniques are used to enhance the functioning of structure level matching by integrating knowledge between data relationships to form a stronger concept base for performing data linkage [75]. Knowledge integration enhances query formulation when the information structure and data sources are not known, as highlighted in [76], and is becoming increasingly important in data matching processes as various data structures conceptualise the same concept in different ways, with resulting inconsistencies and overlapping material. Integration can be based on extensions or concepts, and is aimed at indemnifying inconsistencies and mismatches in the concepts. For example, the COIN technique [77] addresses data-level heterogeneities among data sources expressed in terms of context axioms and provides a comprehensive approach to knowledge integration. An extension of COIN is ECOIN, which improves upon COIN through its ability to handle both data-level and ontological heterogeneities in a single framework [77]. Knowledge integration is highly useful in medicine, to integrate concepts and information within various medical data sources. Knowledge integration involves the introduction of a dictionary to fill knowledge gaps, such as using distance-based weight measurement through Google [68]. For example, the Foundational Model of Anatomy is used as a concept roadmap to better integrate various medical data sources into unique anatomy concepts [68].

## 4.17. Data structures

Data structures use structural information to identify match and reflect relationships. Information properties are often considered and compared with concepts to make a similarity determination, while other variations of the data structure approach uses graphical information to create similarities [68]. A drawback of the data structure based approach results from its consumption rate of resources; the process builds an "in-memory" graph containing paired concepts which can lead to memory overflow.

## 4.18. Statistical analysis

Statistical analysis techniques examine statistical measurements for determining term and concept relationships. Jaccard Similarity Coefficient [38] is a widely used statistical measurement for comparing terms, which consider the extent of overlap between two vectors. The measurement is the size of the intersection, divided by the size of the union of the vector dimension sets. Considering the corpus, the Jaccard Similarity approach determines a match to be present if there is a high probability for both concepts to be present within the same section. For attribute matching, a match is determined if there is a large amount of overlap between values [38]. For example, CORDS [16] is a statistical matching tool, built upon B-HUNT, which locates statistical correlations and soft functional dependencies. CORDS searches for correlated column pairs through enumerating potentially correlating pairs and pruning unqualified pairs. A chi-squared analysis is performed in order to locate numerical and categorical correlations.

Unfortunately, statistical analysis methods are generally restricted to column pairs, and may not detect correlations where not all subsets have been correlated [1, 18].

# 5. CONCLUSIONS & FUTURE WORK

The data linkage approaches reviewed in this paper represents a variety of linkage techniques using different aspects of data. We discussed practical approaches from two different angles, they are, Attribute level and Structure-level approach. We showed that classification of data into a single order does not provide the necessary flexibility for accurately defining data relationships. Furthermore, we found that the flow of data and their relationships need not be in a fixed direction. This is because, when dealing with variable data sources, same sets of data can be ordered in multiple ways based on the semantics of tables, attributes and tuples. This is critical when performing data linkage. Through our analysis of the status quo we also proved that the research should take a new direction to discover possible data matches, based on its inherent hierarchical semantic similarities as proposed in [109]. This approach is ideal for knowledge based data matching and query answering. We recommend faceted classification to classify data in multiple ways, to source semantic information for accurate data linkage and other data intrinsic tasks. We also recommend, in response to the intricacy of this background research, that the data linkage research community collaborate to benchmark existing data linkage techniques, as it is getting increasingly complicated to convincingly and in a timely manner compare new techniques with existing ones.

# AUTHOR

Dr. Mohammed Gollapalli is an Assistant Professor in the College of Computer Science and Information Technology (CCSIT) at the University of Dammam (UD), Dammam, Kingdom of Saudi Arabia. He graduated his PhD in Information Technology at the University of Queensland (UQ), Brisbane, Australia in 2013 and obtained Masters in Information Technology at Griffith University, Gold Coast, Australia, in 2005. His major areas of research interests and expertise include Data Mining, Knowledge Management, and Quality Control. He is a member of MCP and IEEE.

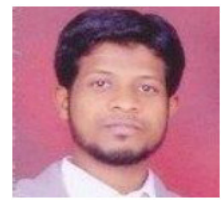